\documentclass[twocolumn,showpacs,preprintnumbers,amsmath,amssymb,superscriptaddress,floatfix]{revtex4}

\usepackage{graphicx}
\usepackage{dcolumn}
\usepackage{bm}
\usepackage{latexsym}
\usepackage{color}

\newcommand{\F}{\phantom {1}}

\newcommand{\Kanazawa}{\affiliation{Institute for Theoretical Physics,
Kanazawa University, Kanazawa 920-1192, Japan}}
\newcommand{\RIKEN}{\affiliation{RIKEN, Radiation Laboratory, Wako 351-0158, Japan}}
\newcommand{\Mainz}{\affiliation{Institute f\"{u}r Kernphysik, Universit\"{a}t Mainz, D-55099 Mainz, Germany}}

\begin{document}

\preprint{KANAZAWA 06-19}
\preprint{MKPH-T-07-01}

\title{Abelian dominance and the dual Meissner 
effect in local unitary gauges \\
in SU(2) gluodynamics}

\author{Toru Sekido}
\Kanazawa
\RIKEN
\author{Katsuya Ishiguro}
\Kanazawa
\RIKEN
\author{Yoshiaki Koma}
\Mainz
\author{Yoshihiro Mori} 
\Kanazawa
\RIKEN
\author{Tsuneo Suzuki}
\Kanazawa
\RIKEN

\date{\today}

\begin{abstract} 

Performing highly precise Monte-Carlo simulations of SU(2) gluodynamics,
we  observe for the first time Abelian dominance in the 
confining part of the static potential in local unitary gauges 
such as the F12 gauge.
We also study the flux-tube profile between the quark and antiquark
in these local unitary gauges and find a clear signal of 
the dual Meissner effect.
The Abelian electric field is found to be squeezed into
a flux tube by the monopole supercurrent.
This feature is the same as that observed in the non-local 
maximally Abelian gauge. 
These results suggest that the Abelian confinement scenario 
is gauge independent. 
Observing the important role of space-like monopoles 
in the Polyakov gauge also indicates that the monopoles defined 
on the lattice do not necessarily correspond 
to those proposed by 't~Hooft in the context of Abelian projection.

\end{abstract}


\maketitle


\par
Quark confinement phenomenon remains an important 
unsolved problem
in quantum chromodynamics (QCD)~\cite{CMI:2000mp}.
One of the most intriguing conjecture for its 
mechanism is that the QCD vacuum behaves as 
a dual superconductor due to magnetic monopole 
condensation~\cite{tHooft:1975pu,Mandelstam:1974pi}, i.e.,
the color flux between a quark and an antiquark
is squeezed into a stringlike tube as the Abrikosov 
vortex~\cite{Abrikosov:1956sx,Nielsen:1973cs}
through the dual Meissner effect, which
yields a linear-confining potential.
Although it is not straightforward to identify the 
corresponding monopoles in QCD
in contrast to SUSY QCD~\cite{Seiberg:1994rs} 
or the Georgi-Glashow model~\cite{'tHooft:1974qc,Polyakov:1976fu}
with scalar fields,
it is possible to reduce SU(3) QCD into an Abelian
[U(1)]$^2$ theory with magnetic monopoles
by a partial gauge fixing, also referred to as
the Abelian projection~\cite{tHooft:1981ht}, and to
accommodate the above dual superconductor 
scenario~\cite{Ezawa:1982bf}.

\par
However, there are infinite ways of the partial gauge-fixing.
Numerically, an Abelian projection with non-local 
gauges such as the maximally Abelian (MA)
gauge~\cite{Suzuki:1983cg,Kronfeld:1987ri,Kronfeld:1987vd} 
has been found to support the Abelian confinement scenario
beautifully~~\cite{Suzuki:1992rw,Chernodub:1997ay,Suzuki:1998hc,Singh:1993jj}.
On the other hand, 
the Abelian confinement mechanism has not been observed clearly 
so far for years in other general gauges
in particular, in local unitary 
gauges~\cite{Suzuki:1989gp,Bernstein:1996vr,Ito:2002zv}.
This is very unsatisfactory, 
since the quark confinement mechanism should
not depend on a special gauge choice~\cite{Carmona:2001ja}.

\par
It is the purpose of this letter to show for the first time 
that the Abelian confinement mechanism is observed 
numerically also in local unitary gauges 
with the method of highly precise numerical simulations.
For numerical simplicity we adopt SU(2) group
instead of SU(3), but the essential feature
of non-Abelian gauge theory should be the same.
As local unitary gauges, we adopt 
simplest candidates, namely the F12, the F123 and 
the spatial Polyakov loop (SPL) gauges as well as 
the original Polyakov (PL) gauge.
Applying the multi-level noise reduction
method invented by L\"{u}scher and Weisz~\cite{Luscher:2001up}, 
we investigate the Abelian static 
potential with high accuracy and find a clear signal of 
Abelian dominance in its confining part 
in all local unitary gauges considered.
Note that F12 (F123) gauge and PL (SPL) gauge are most typical 
but are of completely different types. 
Since we obtain the same results in these different
unitary gauges, we expect that the same can be seen also in other 
local unitary gauges.
In addition, we study the flux-tube profile between 
the quark and antiquark
with the vacuum ensemble
composed of as many as 4000 thermalized configurations 
generated by means of 
the improved Iwasaki action~\cite{Iwasaki:1985we}
and observe the squeezing of Abelian electric 
field into a flux tube due to the magnetic monopole current.
This is the dual Meissner effect and the feature is quite
similar to that already observed in the MA gauge.
The authors expect that the results 
obtained here are very interesting to general readers, since 
they strongly suggest that the Abelian dual Meissner effect caused 
by Abelian monopoles is gauge independent and a correct confinement mechanism.

\begin{figure}[b]
\vspace{-0.8cm}
\includegraphics[height=5.7cm]{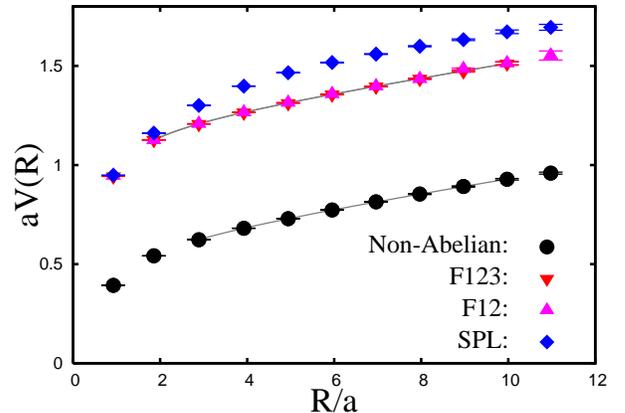}
\vspace{-0.4cm}
\caption{\label{fig-1}
Abelian static potentials 
in the F123, the F12 and the SPL gauges 
in comparison with the non-Abelian potential.
The solid lines denote the best fitting curves to the function 
$V_{\rm fit}(R)$.}
\vspace{-0.3cm}
\end{figure}

\begin{figure}[th]
 \includegraphics[height=5.7cm]{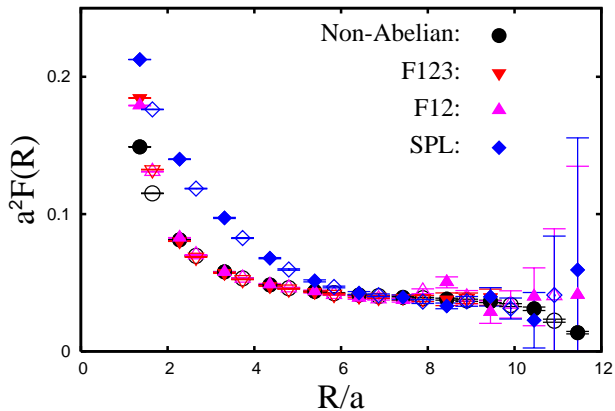}
\vspace{-0.4cm}
\caption{\label{fig-1f}
The force in the F123, the F12 and the SPL gauges 
in comparison with the non-Abelian one.
The filled (open) symbols are forces which 
are defined by using the backward (mid-point) difference.}
\vspace{-0.3cm}
\end{figure}


\par
We generate thermalized gluon configurations 
using the standard SU(2) Wilson gauge action 
at the coupling constant $\beta=2.5$ on the lattice $N^4=24^4$, 
where the lattice spacing is $a(\beta)=0.0836(8)$~fm. 
The scale is set from the string tension
$\sqrt{\sigma}=440$~MeV. 
Periodic boundary conditions are imposed in all directions.
Then we perform a partial gauge fixing diagonalizing  
a plaquette operator 
$X(s) \equiv U_{12}(s)$ [F12 gauge], or
an operator $X(s)\equiv U_{1}(s)U_{2}(s+\hat{1})
U_{3}(s+\hat{1}+\hat{2})U^{\dag}_{1}(s+\hat{3}+\hat{2})
U^{\dag}_{2}(s+\hat{3})U^{\dag}_{3}(s)$ [F123 gauge],
or a space-like Polyakov loop 
$X(s)\equiv 
\mathcal{P}_{\rm SU(2)} 
\sum_{k=1}^3 \prod_{i=0}^{N-1}U_{k}(s+i\hat{k})$
[SPL gauge],
or a usual time-like Polyakov loop 
$X(s)\equiv \prod_{i=0}^{N-1}U_{4}(s+i\hat{4})$ [PL gauge].
After gauge fixing, we decompose SU(2) link variables as
\begin{eqnarray*}
U_{\mu}(s) = U^0_{\mu}(s)
\! + \! 
i\vec{\sigma}\vec{U}_{\mu}(s)= C_{\mu}(s)
\! \left(
 \begin{array}{cc}
    e^{i\theta_{\mu}(s)}   &  0  \\
    0   &   e^{-i\theta_{\mu}(s)} \\
  \end{array}
\right) ,
\end{eqnarray*}
where $\vec{\sigma}$ is the Pauli matrix, and
extract Abelian link variables
$\theta_{\mu}(s)=\arctan
(U^3_{\mu}(s)/U^0_{\mu}(s))$~\cite{Kronfeld:1987vd}.
Since the first three gauges contain only space-like link variables, 
we may apply the multi-level algorithm~\cite{Luscher:2001up} 
to evaluate the Abelian static potential from the correlator of 
the Abelian Polyakov loop defined by
$P_A(\vec{s}) = \exp\left[i\sum_{i=0}^{N-1}\theta_4(s+i\hat{4})\right]$.
For the multi-level method, we choose
the temporal extent of a sublattice to be~$4a$.

Note that the expectation value of such an U(1) invariant
Abelian quantity in some gauges 
is expressed theoretically by a sum of complicated 
gauge-invariant quantities composed of operators 
in various representations \cite{Ogilvie:1998wu} 
and hence the numerical results are not predictable.

\begin{table}[b]
\begin{center}
\vspace{-0.5cm}
\caption{\label{stringtension}
The string tension $\sigma$, the
Coulombic coefficient $c$, and the
constant $\mu$ obtained by the best fit.
FR means the fitting range before tree-level improvement.
$\chi^2$ is defined by the diagonal components of the covariance
matrix and $N_{\rm df}$ is the number of degrees of freedom.
$N_{\rm iupd}$ is the number of internal updates used in 
the multi-level method. 
$N_{\rm conf}$  is $8$.
The errors are estimated by the jackknife method.}
\begin{tabular}{c|cccccc}
  & $\sigma a^2$ & $c$     & $\mu a$  & FR(R/a)& $\chi^2/N_{\rm df}$ & 
     $N_{\rm iupd}$ \\ \hline
NA   & .0348(7)  & .243(6) & 0.607(4) & 3 - 10 & 0.35  & 15000 \\
F123 & .0350(2)  & .239(1) & 1.187(1) & 2 - 10 & 0.10  & 80000 \\
F12  & .0345(6)  & .244(4) & 1.192(3) & 2 - 10 & 1.08  & 80000 \\
\end{tabular}
\end{center}\vspace{-0.7cm}
\end{table}

\begin{figure}[hbt]
\includegraphics[height=5.cm]{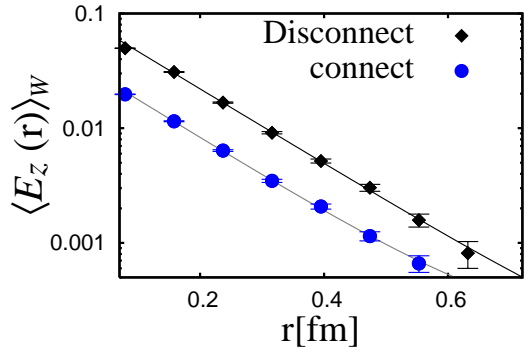}\\
\vspace{-0.5cm}
\caption{\label{fig-3} The profile of the Abelian electric field 
from the connected
and disconnected correlators in the MA gauge.}
\vspace{-0.5cm}
\end{figure}

\par
In Fig.~\ref{fig-1}, we show the Abelian static potential
as well as the non-Abelian potential as 
a function of the $q$-$\bar{q}$ distance~$R$,
where a tree-level perturbative improvement of the
distance is applied to avoid an enhancement of lattice artifacts 
especially at short distances~\cite{Necco:2001xg,Luscher:2002qv}. 
We find that the results in local unitary gauges 
are remarkably clean. 
Note that although the number of configurations ($N_{\textrm{conf}}$) for 
measurements is low, the multi-level algorithm with enough internal updates gives us clear expectation values of the Abelian Polyakov-loop correlators up to as precisely as $10^{-16}$.
The slope at large distances looks identical.
We may fit the potentials to a usual functional form
$V_{\rm fit}(R)=\sigma R - c/R + \mu$ and
extract the string tension $\sigma$.
The best fitting parameters
are summarized in Table~\ref{stringtension}.
In the F12 and the F123 gauges, 
the string tensions 
are in almost complete agreement with the non-Abelian one.
 In the SPL gauge ($N_{\rm conf}=10$ 
with $N_{\rm iupd}=300000$), 
however, the Coulombic coefficient becomes so large 
($c\approx 0.8$) that we cannot determine the string tension
definitely on this small lattice.
On the other hand, in all gauges the force,
which is defined by differentiating the potential with respect to $R$,
shows a good agreement at large distances (see Fig.~\ref{fig-1f}).
In the PL gauge, the agreement of Abelian and non-Abelian string 
tensions is trivial, since non-Abelian Polyakov lines are equal to 
Abelian ones in this gauge.


\par
Next let us study the dual Meissner effect in local unitary gauges. 
In this calculation, we employ the improved Iwasaki gauge 
action with the coupling constant 
$\beta= 1.20$,
which corresponds to the lattice spacing 
$a(\beta)= 0.0792(2)$~fm~\cite{Suzuki:2004dw}.
The lattice size is $32^4$ with periodic boundary conditions. 
We have taken 4000 thermalized configurations for measurements. 
To improve a signal-to-noise ratio, 
the APE smearing technique is 
applied to the Wilson loop~\cite{Albanese:1987ds}. 
In addition, although we could measure usual (disconnected) correlators
after the gauge-fixing,we evaluate a connected correlator defined by
$\langle {\cal O}_A(r) \rangle_{W}
=
\langle \mbox{Tr}\left[
LW(R,T)L^{\dagger}\sigma^3{\cal O}_A^3(r)
\right]\rangle
/ \langle \mbox{Tr}\left[W(R,T)\right]\rangle$
for various operators ${\cal O}_A$ composed of Abelian link 
variables in order to get a better signal, where
$L$ is a product of non-Abelian link variables called the Schwinger 
line, connecting the Wilson loop $W$ with the Abelian 
operator~\cite{Cea:1995zt, DiGiacomo:1989yp}.
$r$ is the minimal distance between $W$ and ${\cal O}_A$.
As Abelian operators, for instance, we employ the Abelian electric field 
$E_{i}(s)\equiv \bar{\theta}_{4i}(s)$ and 
the monopole current $k_{i}(s)\equiv
(1/4\pi)\epsilon_{i\nu\rho\sigma}\partial_{\nu}
\bar{\theta}_{\rho\sigma}(s+\hat{i})$ 
where $\theta_{\mu\nu}(s) \equiv \theta_{\mu}(s)+\theta_{\nu}(s+\hat{\mu})
-\theta_{\mu}(s+\hat{\nu})-\theta_{\nu}(s)
=\bar{\theta}_{\mu\nu}(s)+2\pi n_{\mu\nu}(s) 
~(|\bar{\theta}_{\mu\nu}|<\pi)$~.

\begin{figure}[t]
 \includegraphics[height=5.cm]{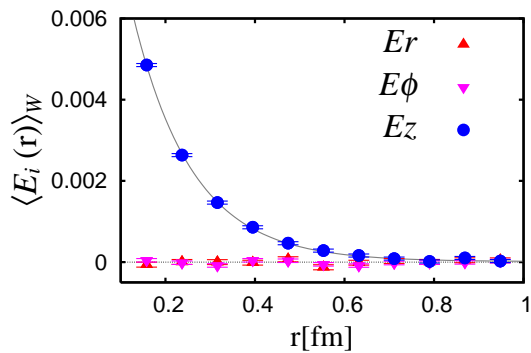}
\vspace{-0.6cm}
\caption{\label{fig-4}
The profile of the Abelian electric field in the F123.
The solid line denotes the fitting curve for
$\langle E_{z}(r)\rangle_{W}$ 
to the function $f(r)$.}
\vspace{-0.4cm}
\end{figure}

\par
We first examine the consistency between the result 
from the usual disconnected
correlators~\cite{Singh:1993jj} 
and the connected one in the MA gauge 
for $\langle E_{z}(r) \rangle_{W}$. 
The profile of the electric field is plotted in Fig.~\ref{fig-3}.
We find that while the absolute magnitude of the electric field 
depends on the type of the correlator, their exponential decay
rates look the same.
Assuming a functional form $f(r)=c_1 \exp(-r/\lambda)+c_0$,
we may estimate the penetration length $\lambda$,
which characterizes the strength of the dual Meissner effect.
The result is $\lambda=0.133(4)$~fm for the disconnected correlator
and $\lambda=0.131(10)$~fm for the connected one
in the case of $W(5a,7a)$. 
$c_0$ is consistent with zero in each case.
This  indicates that the result of the
connected correlator can be consistent with that of disconnected one.

\par
In Fig.~\ref{fig-4}, 
we show the Abelian electric field profile in 
the F123 gauge for $W(5a,5a)$.
We find that only $\langle E_{z}(r) \rangle_{W}$ 
exhibits an exponential decay as a function of $r$ and the 
penetration length is then found to be $\lambda=0.133(3)$~fm.
As seen from Table~\ref{penetration},
$\lambda$ in other unitary gauges are almost the same, 
which are also consistent with that in the 
MA gauge~\cite{Chernodub:2005gz}. 
Note that we also investigate the profile of the magnetic field
with the operator 
$B_{i}(s)=(1/2)\epsilon_{ijk}\bar{\theta}_{jk}(s)$
and find no correlation with the Wilson loop.

\begin{table}[t]
\caption{\label{penetration}
The penetration length $\lambda$ and the 
coherence length $\xi$  both in unit of~fm 
($\kappa \equiv \lambda/\xi$),
which are evaluated from $W(5a,5a)$.
MA(d) means the disconnected correlator in the MA gauge.
In each case  $\chi^2/N_{\rm df}=0.6 - 1.1$.
}
\begin{tabular}{c|cccccc}
           & MA(d)    & F123     & F12      & SPL          & PL \\
\hline
 $\lambda$ & 0.129(2) & 0.133(3) & 0.132(3) & 0.134(7){\F} & 0.132(4){\F} \\
 $\xi$     & 0.154(6) & ---      & ---      & 0.162(38)    & 0.142(28) \\
\hline
 $\sqrt{2}\kappa$ & 1.18(6) & --- & ---     & 1.17(34)     & 1.31(29)
\end{tabular}
\vspace{-0.4cm}
\end{table}

\begin{figure}[b]
\vspace{-0.6cm}
 \includegraphics[height=5.4cm]{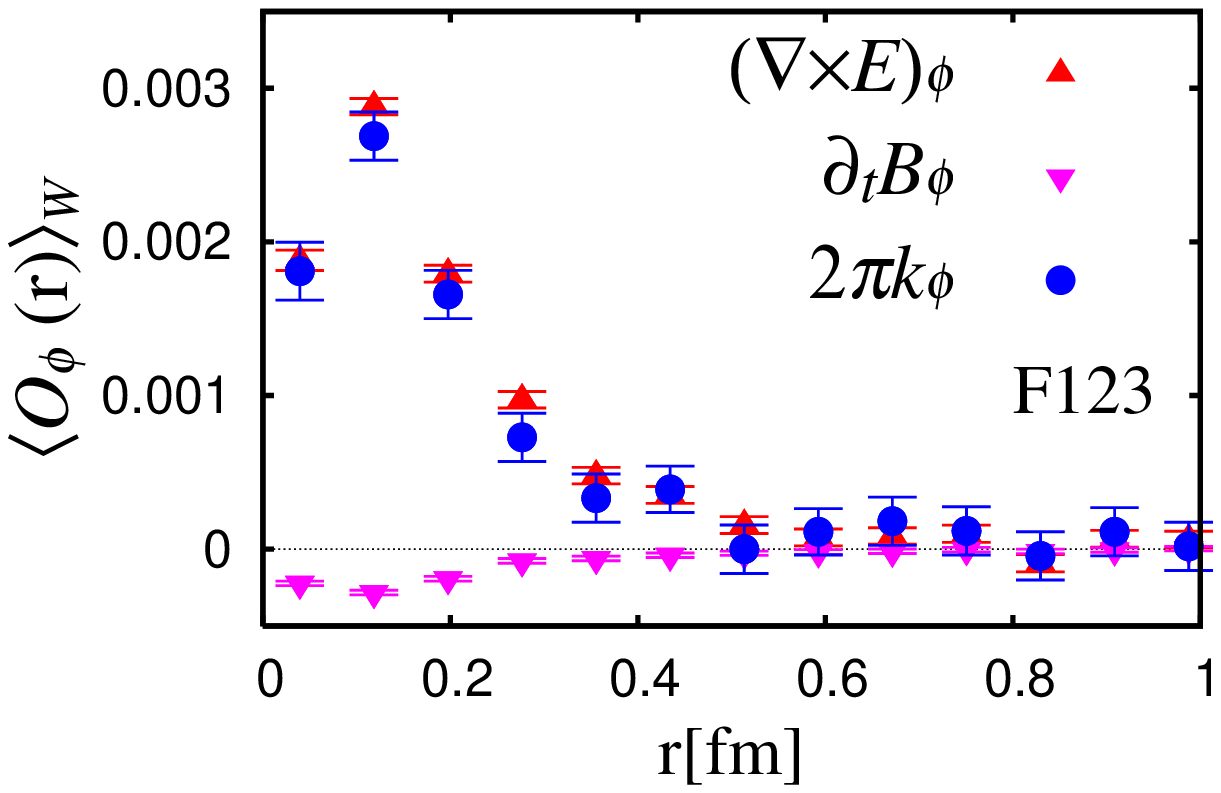}\\
\vspace{-0.4cm}
 \includegraphics[height=5.4cm]{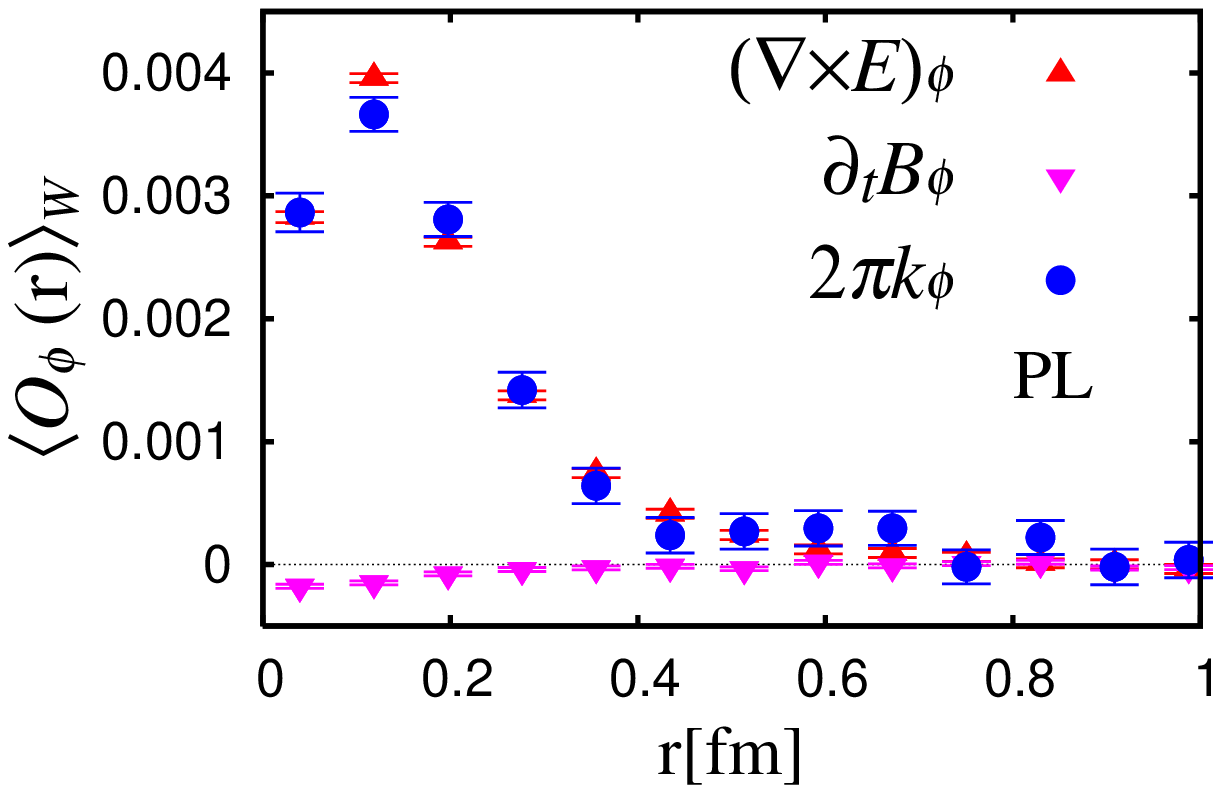}
\vspace{-0.5cm}
\caption{\label{fig-5} Curl of the Abelian electric field,
the magnetic
displacement current and the monopole current 
in the F123 (upper) and PL (lower) gauges.}
\end{figure}

\par
To identify what squeezes the Abelian electric field, 
let us study 
the Abelian (dual) Amp\`ere law
$\vec{\nabla}\times\vec{E}=\partial_{t}\vec{B}+2\pi\vec{k}$~.
In Fig.~\ref{fig-5},  we show the profile of each term 
in the F123 and the PL gauges, where only the 
non-vanishing azimuthal components are plotted.
We find that the curl of electric field 
$\vec{\nabla}\times\vec{E}$
is reproduced by the monopole current 
$2\pi\vec{k}$, while the magnetic displacement 
current $\partial_{t}\vec{B}$ gives only small contribution.
The magnitude of the profile depends on the gauge, but the 
qualitative feature is quite similar in both gauges.
Note that this behavior is consistent with that in the MA 
gauge~\cite{Suzuki:2004dw}.

\par
It is important to note that the space-like monopole current
is responsible for squeezing the electric field in the PL gauge. 
This suggests that the monopoles defined on the lattice 
do not necessarily correspond to 't~Hooft's Abelian 
monopoles~\cite{tHooft:1981ht}, where the latter is due to 
the degeneracy points of eigenvalues of some 
adjoint operators and they are always time-like 
in the PL gauge~\cite{Chernodub:2003mm}.
Rather, if the monopoles we observe on the lattice do not 
exactly correspond to the 't~Hooft  monopoles, the role of
monopoles for the Abelian confinement mechanism can be
gauge independent, and indeed, the above results 
seem to support such an expectation.

\begin{figure}[t]
 \includegraphics[height=5.3cm]{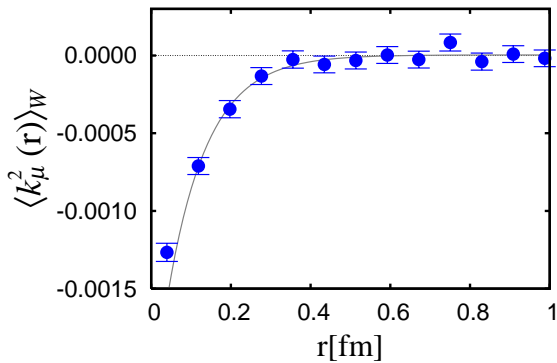}
\vspace{-0.6cm}
\caption{\label{fig-7} The profile of
the squared monopole density $\langle k_{\mu}^2(r) \rangle_{W}$ 
in  the PL gauge for $W(5a,5a)$.
The solid line denotes the fitting curve to the function $g(r)$.}
\vspace{-0.4cm}
\end{figure}

\par
Finally, let us briefly discuss the type of the dual 
superconductor vacuum.
For this purpose it is important to evaluate the 
coherence length $\xi$
as well as the penetration length $\lambda$.
The ratio of these two length scales, the GL parameter
$\kappa = \lambda/\xi$, classifies the vacuum type.
$\sqrt{2}\kappa = 1$ 
corresponds to the border
between the type~I and the type~II vacua.
As demonstrated in Ref.~\cite{Chernodub:2005gz} we may 
extract~$\xi$
by fitting the profile of the squared monopole density
$\langle k_{\mu}^2(r) \rangle_{W}$
to a functional form 
$g(r)=c'_1\exp(-\sqrt{2} r /\xi)+c'_0$.
For the operator $k_{\mu}^2$ 
the connected correlator is reduced to the 
disconnected one and here we evaluate the latter.
In Fig.~\ref{fig-7}, we show the profile in the
PL gauge for $W(5a,5a)$.
We obtain a similar behavior in the SPL gauge.
However, we cannot identify the profile 
in the F12 and the F123 gauges within statistics,
which is probably  due to contamination from many
ultraviolet monopoles.
In Table~\ref{penetration}, we summarize the coherence length 
in the SPL and the PL gauges as well as that in the MA gauge 
for comparison.
The value of the GL parameter looks consistent with each other.
However, we note that the value may not be definite
since the Wilson loop size adopted here is still small and
further systematic studies are required.
What we can conclude here is that the vacuum type 
of SU(2) gluodynamics is not far from the border.


\par
In conclusion, we have observed
a clear signal of Abelian dominance in the confining
part of the static potential in local unitary gauges
for the first time.
The structure of the flux-tube profile in these gauges
strongly support a gauge-independent
role of monopoles in the Abelian confinement scenario.
The study of Abelian confinement mechanism
without performing any gauge fixing
is in progress and the result will be published elsewhere.

\par
The authors
thank to M.~Polikarpov, V.~Zakharov, M.~Chernodub, V.~Bornyakov 
and G. Schierholz for fruitful discussions. 
The numerical simulations
of this work were done using RSCC computer clusters in 
RIKEN and SX5 at RCNP of Osaka University. The authors 
would like to thank RIKEN and RCNP for their support of 
computer facilities. 
T.S. is supported by JSPS Grant-in-Aid for Scientific 
Research on Priority Areas 13135210.

\bibliographystyle{ref/h-physrev3}
\bibliography{ref/confine,ref/suzuki-mod,ref/maxim-mod,ref/fedor}

\end{document}